\documentclass[aps, prb, twocolumn, groupedaddress, showpacs, floatfix]{revtex4-1}
\usepackage{amsmath}
\usepackage{amssymb}
\usepackage{graphicx}
\usepackage{xr-hyper}
\usepackage{bm}

\begin{document}
\title{Scattering Theory Approach to Inelastic Transport in Nanoscale Systems}

\author{Sejoong Kim}
\email[ ]{sejoong@kias.re.kr}
\author{Young-Woo Son}
\affiliation{Korea Institute for Advanced Study, Seoul 130-722, Korea}

\date{\today}

\begin{abstract}
We present a scattering-state description for the non-equilibrium multichannel charge transport in the presence of electron-vibration couplings. It is based on an expansion of scattering orders of eigenchannel states. Examining charge transitions between scattering states, we clarifies competing inelastic and elastic scattering processes, and compare with the interpretation based on the non-equilibrium Green's functions formalism. We also derive a general expression for conductance variations in single-channel systems. It provides a comprehensive picture for the variation including the well-known result, the 0.5 rule, from the aspect of interplay between elastic and inelastic scattering processes. 
\end{abstract}


\maketitle

\section{Introduction}
Understanding interactions between conducting electrons and molecular vibrations is of central importance in analyzing spectra from inelastic electron tunneling spectroscopy (IETS) experiments~\cite{PRL1966Lambe, Science1998Stipe, Nature2002Smit, PRL2002Agrait} and is one of critical issues in developing the future molecular electronics~\cite{JPhysCM2007Galperin}. 
IETS signals are characterized by the vibration-induced conductance variations at a threshold bias voltage equal to a vibrational energy, which indicate opening of inelastic channels~\cite{PRL1966Lambe, Science1998Stipe, Nature2002Smit, PRL2002Agrait}. 
The conductance variation undergoes a crossover from an increase to a decrease when a system evolves from a low-conductance regime to a high-conductance one~\cite{PRL2008Tal, PRB2006Vega, PRB2005Paulsson, PRL2008Paulsson}. In particular, it has been argued that the crossover occurs when a bare transmission is approximately a half, which is called as the $0.5$ rule~\cite{PRL2008Tal, PRB2006Vega, PRB2005Paulsson, PRL2008Paulsson}. A typical system exhibiting the $0.5$ crossover is a single-level model symmetrically coupled to electrodes~\cite{PRL2008Paulsson, PRB2007Frederiksen_b}. However, many cases are not simply explained by this rule~\cite{PRB2007Frederiksen_b, PRB2012Avriller, PRL2000Hahn, PRB2009Kristensen}. When the single-level model is not symmetrically connected to electrodes, then the crossover deviates from $0.5$~\cite{PRL2008Paulsson}. Even in symmetric junctions, odd symmetric vibrational modes do not lead to the conductance crossover~\cite{PRB2007Frederiksen_b, PRB2012Avriller}. In the scanning tunneling microscope experiment~\cite{PRL2000Hahn}, it is reported that a stretching vibration of an oxygen molecule on a silver substrate gives rise to a conductance decrease in a low-conductance regime. Furthermore, multichannel systems can show a co-ocurrence of positive and negative conductance steps~\cite{PRB2009Kristensen, arXiv2012Sejoong}. Thus it is required to comprehensively understand the conductance variation and related physics in a single framework.

When electrons interact with local vibrations, their energy can either change or not, depending on scattering processes. When electrons emit or absorb energy during scattering with vibrations, it is defined as the inelastic scattering process. On the other hand, the energy of conducting electrons can be conserved via emission-reaborption or absorption-reemission processes. This is termed the elastic scattering process. The conductance variation is determined by a competition between elastic and inelastic scattering channels~\cite{PRB1970Davis, JPhysC1972Caroli, PRL1987Persson, PRL2000Lorente, PRB2003Mii, JChP2004Galperin, PRB2005Viljas, SurfSci2007Ueba}. Specifying those scattering processes and their interplay is a key factor to understand all of the reported observations. Although some works investigate the crossover of conductance steps in specific systems such as the single-level model~\cite{PRL2008Paulsson} and atomic chains~\cite{PRB2006Vega, PRB2007Frederiksen_b}, it has not been clearly discussed how the competition of the two scattering processes leads to the crossover in general situations. 

In this paper, we present a scattering theory description of inelastic electron transport, emphasizing the role of inelastic and elastic scattering processes on the conductance variation. Our description can be made possible by converting non-equilibrium Green's functions (NEGFs) to scattering states. Among various inelastic transport theories~\cite{PRB2006Vega, PRB2005Paulsson, PRL2008Paulsson, PRB2007Frederiksen_b, PRB2012Avriller, PRL2000Hahn, PRB2009Kristensen, arXiv2012Sejoong, PRB1970Davis, JPhysC1972Caroli, PRL1987Persson, PRL2000Lorente, PRB2003Mii, JChP2004Galperin, PRB2005Viljas, SurfSci2007Ueba, PRL1995Bonca, PRL1999Ness, PRB2000Emberly, PRL2001Lorente, 2003Montgomery1, 2003Montgomery2, NanoLett2003Chen, NanoLett2004Chen, NanoLett2005Chen, NanoLett2005Jiang, PRB2005Troisi, PRL2004Frederiksen, PRB2007Frederiksen_a, NanoLett2004Pecchia, PRB2007Pecchia, PRB2010Romano}, the NEGF theory~\cite{PRB2006Vega, PRB2005Paulsson, PRL2008Paulsson, PRB2007Frederiksen_b, PRB2012Avriller, PRL2000Hahn, PRB2009Kristensen, arXiv2012Sejoong, JPhysC1972Caroli, PRL2000Lorente, PRB2003Mii, JChP2004Galperin, PRB2005Viljas, SurfSci2007Ueba, PRL2004Frederiksen, PRB2007Frederiksen_a, NanoLett2004Pecchia, PRB2007Pecchia, PRB2010Romano, PRB2007Gagliardi} has been widely used to calculate inelastic transport properties for realistic IETS setups, implemented with first-principle methods. In order to understand results from the NEGF theory, it is needed to use {\it a posteriori} analysis based on approximate scattering states~\cite{PRB2009Kristensen, PRL2008Paulsson, JChemPhys2006Troisi, PRB2007Gagliardi}. However, some of those analyses do not take into account both elastic and inelastic scattering contributions~\cite{PRB2009Kristensen, PRB2007Gagliardi}. In addition, Ref.~\onlinecite{PRL2008Paulsson} proposed an ansatz based on scattering rates in the form of Fermi's golden rule, instead of directly dealing with the interplay of elastic and inelastic contributions. It seems that the ansatz is reasonable to qualitatively examine the experiemtnal selection rule known as {\it propensity rule}~\cite{PRL2008Paulsson, JChemPhys2006Troisi, PRB2007Gagliardi}, that only some of the vibrational modes contribute to inelastic signals. 
The ansatz, however, does not predict signs of conductance steps, and it is questionable that it can quantitatlvely reproduce heights of conductance variations. Furthermore, many scattering-state approaches~\cite{2003Montgomery1, 2003Montgomery2, NanoLett2003Chen, NanoLett2004Chen, NanoLett2005Chen} developed independently of the NEGF formulation rely only on the first Born approximation of a scattering theory, which corresponds to the inelastic scattering process. 

In our description, the scattering processes can be expressed within the first and second Born approximations clarifying elastic and inelastic contributions to the current correction.
In this regard, our result is a generalization of Ref.~\onlinecite{PRB1970Davis}, in which the first and second Born approximations are treated on an equal footing. While Ref.~\onlinecite{PRB1970Davis} is limited to a one-dimensional square-well potential with a single vibrational scatterer, our result can explain the conductance variation in general situations involving many vibrational modes as well as multiple scattering channels, without relying on any particular system or any ansatz. 

Considering single-channel systems, we establish an expression for the conductance jumps in the form of Fermi's golden rule. We find that the elastic scattering process is negative, while
the sign of the inelastic contribution changes when a bare transmission is $0.5$. When combining the two contributions, the crossover between postive and negative conductance steps can generally occur at a bare transmission smaller than $0.5$, or even the crossover does not take place. The $0.5$ crossover is recovered when the elastic contribution is suppressed. 

This paper is organized as follows. In Sec.~\ref{theory} we summarize the NEGF theory for inelastic transport, and we derive the scattering-state description from the NEGF formulation. In Sec.~\ref{interpretation}, we identify elastic and inelastic processes in our scattering theory expression by focusing charge transfers between scattering states and energy exchange between conducting electrons and local vibrations. We compare this identification with one used in many references~\cite{PRB2006Vega, PRL2008Paulsson, JPhysC1972Caroli, PRL2000Lorente, PRB2003Mii, JChP2004Galperin, PRB2005Viljas, SurfSci2007Ueba}. In Sec.~\ref{discussion}, we apply our result to single-channel systems. After discussing crossover transmissions for general cases, we consider mirror-symmetric systems and the single-level model. We also make a remark on how our expression can be used for multichannel systems. In Sec.~\ref{conclusion} we make a final conclusion. Technical details and derivations are discussed in Appendices~\ref{Fermi_factors} and \ref{derivation}.

\section{Theory}\label{theory}
\subsection{Non-Equilibrium Green's Functions}
We start with the non-equilibrium transport theory based on NEGFs in a weak electron-vibration ({\it el-vib}) coupling regime. 
In a zero temperature limit and a regime where a damping rate of vibrations is much larger than a heating rate, 
the NEGF formalism gives the current correction $\delta I = \delta I_{1} + \delta I_{2}$ leading to conductance steps 
when a bias voltage of $eV$ is equal to a vibrational energy $\hbar\omega_{\lambda}$ (see Ref.~\onlinecite{PRB2007Frederiksen_a} or Appendix~\ref{Fermi_factors}):
\begin{eqnarray}
\label{I1}
\delta I_{1} = \frac{2e}{h}\sum_{\lambda}\int_{\mu_{R}+\hbar\omega_{\lambda}}^{\mu_{L}}d\varepsilon\textrm{Tr}\left[G_{0}^{a}\Gamma_{L}G_{0}^{r}\mathcal{M}^{\lambda}A_{R}^{-} \mathcal{M}^{\lambda}\right],
\end{eqnarray}
and
\begin{eqnarray}
\label{I2}
\delta I_{2} &=& \frac{2e}{h} \sum_{\lambda} \int^{\mu_{L}}_{\mu_{R}+\hbar\omega_{\lambda}} d\varepsilon \textrm{Im} \textrm{Tr} \left[ \Gamma_{L} G_{0}^{r} \mathcal{M}^{\lambda} A_{R}^{-} \mathcal{M}^{\lambda} A_{R} \right] \nonumber \\
         & & + \int^{\mu_{L}-\hbar\omega_{\lambda}}_{\mu_{R}} d\varepsilon \textrm{Im} \textrm{Tr} \left[ \Gamma_{L} G_{0}^{r} \mathcal{M}^{\lambda} A_{L}^{+} \mathcal{M}^{\lambda} A_{R} \right].
\end{eqnarray}
Here the lesser and greater {\it el-vib} self-energies $\Sigma^{\lessgtr,(2)}_{el-vib}$ give $\delta I_{1}$, while $\delta I_{2}$ originates from the retarded and advanced {\it el-vib} self-energies $\Sigma_{el-vib}^{r,a}$. 
$\mathcal{M}^{\lambda}$ is the {\it el-vib} interaction for the vibrational mode $\lambda$. 
$G^{r(a)}_{0}$ is the retarded (advanced) Green's function of the conductor part without {\it el-vib} interactions,
\begin{equation}
G^{r(a)}_{0} = \left[\varepsilon-\mathcal{H}_{C}-\Sigma_{\textrm{lead}}^{r(a)}\right]^{-1},
\end{equation}
where $\mathcal{H}_{C}$ is the conductor Hamiltonian, and $\Sigma^{r(a)}_{\textrm{lead}}$ is the retarded (advanced) lead self-energy~\cite{1995Datta}. 
$\Gamma_{\alpha}$ is the coupling function to leads, and $A_{\alpha} = G_{0}^{r} \Gamma_{\alpha} G^{a}_{0}$ is the spectral function of the conductor region originating from the electrode $\alpha$ $(=L,R)$. 
Superscripts $\pm$ indicate that energy argument is $\varepsilon \pm \hbar \omega_{\lambda}$ and without $\pm$, the argument is $\varepsilon$. 
Here we assume that the left chemical potential $\mu_{L}$ is bigger than the right one $\mu_{R}$.

\subsection{Scattering theory}  
Our scattering-state description is established by converting NEGFs to scattering states explicitly. 
It can also be done by directly applying scattering theory and taking into account the electron statistics properly, as done in Ref.~\onlinecite{PRB1970Davis}. 
Here, by using a relationship between NEGFs and scattering states, we can make a direct comparison between the two theories, 
and discuss how differently scattering processes are interpreted.
For scattering states, we choose the transmission eigenchannel representation~\cite{PRB1992Martin, PRB2001Lee, PRB2007Paulsson}, in which the scattering matrix $\mathcal{S}$ can be decomposed into 
a collection of $2\times2$ block scattering matrices $\mathcal{S}_{m}=\left(\begin{smallmatrix} r_{m} & t_{m}^{\prime}\\ t_{m} & r_{m}^{\prime}\end{smallmatrix}\right)$
for eigenchannel states $\{ |\Phi_{Lm}\rangle, |\Phi_{Rm}\rangle \}$. 
Note that the energy normalization $\langle \Phi_{\alpha m} (\varepsilon) | \Phi_{\beta n} (\varepsilon^\prime) \rangle = \delta_{\alpha \beta} \delta_{m,n} \delta(\varepsilon - \varepsilon^\prime)$ is used for scattering states~\cite{PRB2007Paulsson}. 
Scattering states $\{|\Phi_{\alpha m} \rangle \}$ are related to Green's functions of the conductor part $G^{r}_{0}$ in the following way~\cite{PRB2009Wang}:
\begin{equation}
\label{identity1} |\Phi_{\alpha m}\rangle = \frac{1}{\sqrt{2\pi}} G_{0}^{r} |W_{\alpha m}\rangle,
\end{equation}
where $|W_{\alpha m} \rangle=\sqrt{2\pi} V_{C\alpha} |u_{\alpha m} \rangle$ and $V_{C\alpha}$ is the coupling Hamiltonian between the conductor and the electrode $\alpha$.
$|u_{\alpha m} \rangle$ is the eigenchannel scattering state when $V_{C\alpha} = 0$, which is just a sum of the incident wave and the totally reflected one. 
$\Gamma_{\alpha}$ is written as $\Gamma_{\alpha} = \sum_{m} |W_{\alpha m} \rangle \langle W_{\alpha m} |$. 
Reference~\onlinecite{PRB2009Wang} provides the Fisher-Lee relation~\cite{PRB1981FisherLee} for the scattering matrix $\mathbf{S}$ of scattering states $\{ |\Psi_{\alpha m}\rangle \}$ before the eigenchannel transformation is applied,
\begin{equation}
\mathbf{S}_{\alpha m,\beta n}
\label{FisherLee}=\left(-\delta_{\alpha\beta}\delta_{mn}+i 2\pi \langle \Psi_{\alpha m}| {G_{0}^{r}}^{-1} |\Psi_{\beta n}\rangle\right).
\end{equation}
The scattering matrix $\mathcal{S}$ can be obtained from Eq.~(\ref{FisherLee}) via the eigenchannel transformation~\cite{PRB1992Martin, PRB2001Lee}. 

If the system respects the time-reversal symmetry, one can prove following relations by using Eq. (\ref{identity1}):
\begin{eqnarray}
\label{identity2-1}\hat{\Theta} |\Phi_{Lm}\rangle &=& r_{m}^{*}|\Phi_{Lm}\rangle+t_{m}^{*}|\Phi_{Rm}\rangle\\
\label{identity2-2}\hat{\Theta} |\Phi_{Rm}\rangle &=& t_{m}^{*}|\Phi_{Lm}\rangle+r_{m}^{\prime *}|\Phi_{Rm}\rangle,
\end{eqnarray}
where $\hat{\Theta}$ is the time-reversal operator. Note that these relations are true not only far from the scattering region, but also inside the conductor. 

After plugging Eqs. (\ref{identity1})-(\ref{identity2-2}) into Eqs. (\ref{I1}) and (\ref{I2}) and then reorganizing each term in a resulting equation
according to scattering orders (see Appendix~\ref{derivation} for detailed derivation), we obtain the $\delta I$ that has the two components $\delta I_{1BA}$ and $\delta I_{2BA}$, 
\begin{widetext}
\begin{eqnarray}
\label{inelastic} \delta I_{\textrm{1BA}} &=& \frac{2e}{h} \left(2\pi\right)^{2}\sum_{\lambda}\sum_{m,n}\int_{\mu_{R}+\hbar\omega_{\lambda}}^{\mu_{L}}d\varepsilon \left(\mathcal{R}_{n}^{-}-\mathcal{T}_{m}\right) \left|\langle\Phi_{Rn}^{-}|\mathcal{M}^{\lambda}|\Phi_{Lm}\rangle\right|^{2},
\end{eqnarray}
and
\begin{eqnarray}
\label{elastic} \delta I_{\textrm{2BA}} &=& -\frac{2e}{h}\left(2\pi\right)^{2}\sum_{\lambda}\sum_{m,n}\int_{\mu_{R}+\hbar\omega_{\lambda}}^{\mu_{L}}d\varepsilon \textrm{Re}\left[  r_{m}^{\prime}t_{m}^{*}\langle\Phi_{Rm}|\mathcal{M}^{\lambda}|\Phi_{Rn}^{-}\rangle\langle\Phi_{Rn}^{-}|\mathcal{M}^{\lambda}|\Phi_{Lm}\rangle\right]  \nonumber\\
                                        & &  +\frac{2e}{h}\left(2\pi\right)^{2}\sum_{\lambda}\sum_{m,n}\int_{\mu_{R}}^{\mu_{L}-\hbar\omega_{\lambda}}d\varepsilon \textrm{Re}\left[r_{m}^{\prime}t_{m}^{*}\langle\Phi_{Rm}|\mathcal{M}^{\lambda}|\Phi_{Ln}^{+}\rangle\langle\Phi_{Ln}^{+}|\mathcal{M}^{\lambda}|\Phi_{Lm}\rangle\right],
\end{eqnarray}  
\end{widetext}
where $\mathcal{R}^{-}_{n} = \left|r_{n} \left( \varepsilon-\hbar\omega_{\lambda} \right) \right|^2$ and $\mathcal{T}_{m} = \left|t_{m} \left( \varepsilon \right) \right|^2$.
We note that Eqs. (\ref{inelastic}) and (\ref{elastic}) are not equal to Eqs. (\ref{I1}) and (\ref{I2}) from the NEGF formalism respectively, {\it i.e.}, $\delta I_{1} \neq \delta I_{\textrm{1BA}}$ and $\delta I_{2} \neq \delta I_{\textrm{2BA}}$.

\section{Interpretation}\label{interpretation}
\begin{figure}[t]
\begin{center}
\includegraphics[width=0.9\columnwidth, clip=true]{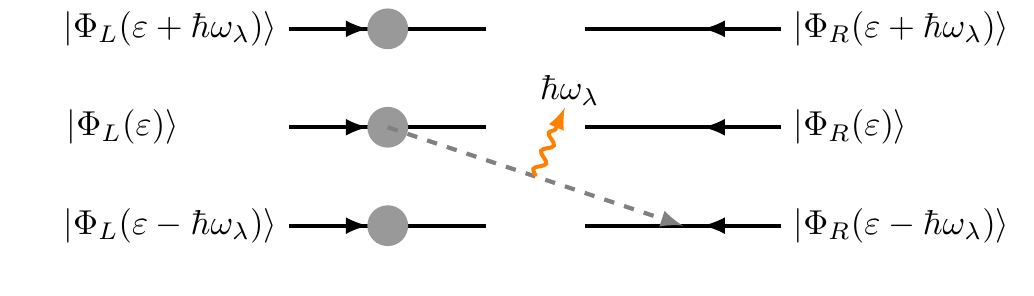} 
\end{center}
\caption{\label{1BA} (Color online) Schematic explanation of scattering processes in Eq.(\ref{inelastic}).  The thick right (left) arrows denote eigenchannels $|\Phi_{L}\rangle$ ($|\Phi_{R}\rangle$). The dotted arrow indicates the inelasitc scattering process emitting a vibron. The orange wiggly line represents a vibron emission.}
\end{figure}

\begin{figure}[t]
\begin{center}
\includegraphics[width=1.0\columnwidth, clip=true]{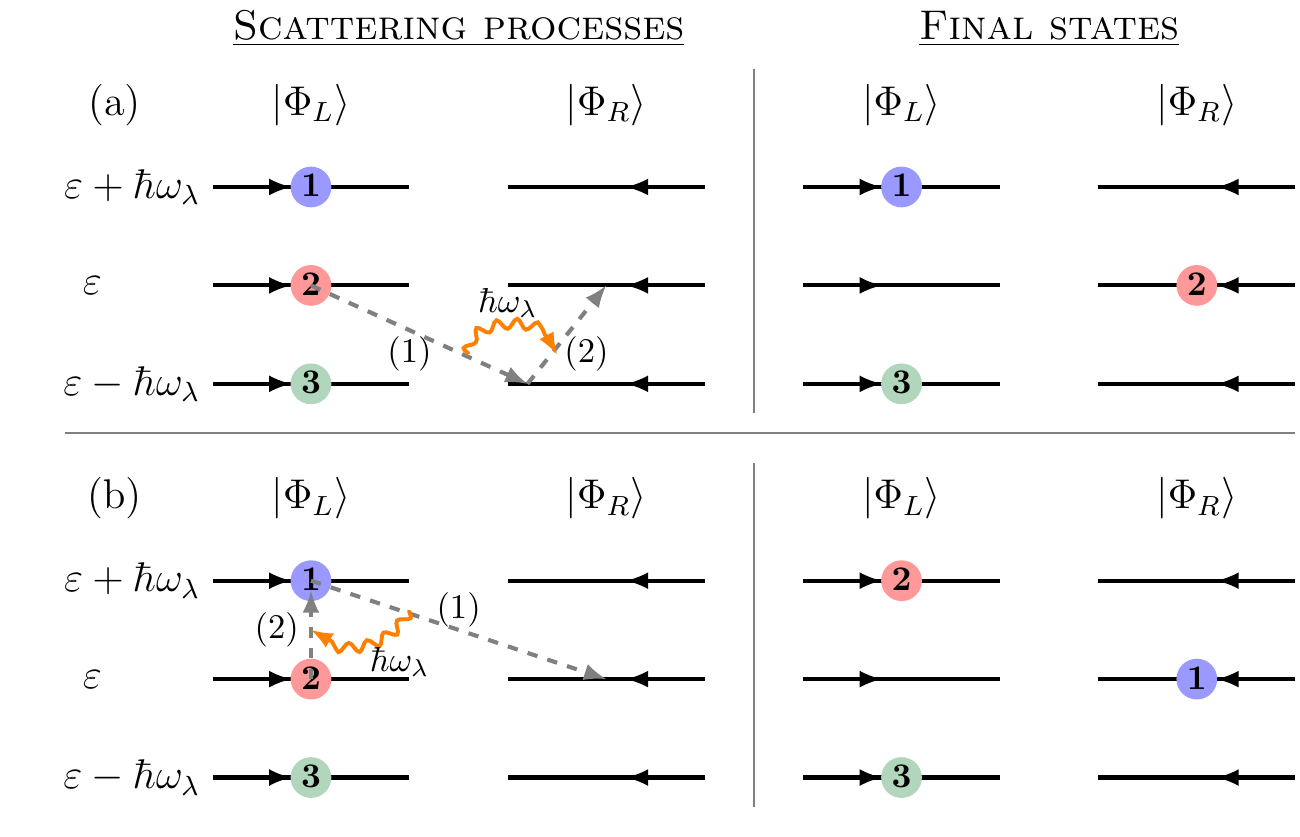} 
\end{center}
\caption{\label{2BA} (Color online) Schematic explanation of scattering processes in the interference term of Eq.(\ref{elastic}), which accompany the vibrational excitation-deexcitation. 
The thick right (left) arrows denote energy levels of electrons corresponding to $\varepsilon + \hbar \omega_{\lambda}$, $\varepsilon$, $\varepsilon - \hbar \omega_{\lambda}$
belong to eigenchannels $|\Phi_{L}\rangle$ ($|\Phi_{R}\rangle$). The dotted arrows associated with numberings $(1)$ and $(2)$ indicate the order of scattering events. Orange wiggly lines represent vibronic energy transfer. 
(a) one-electron scattering process. 
(b) two-electron scattering process. 
The right panel shows the final states of two scattering processes. Note that the final ones in (a) and (b) are equivalent under the exchange of the electrons $1$ and $2$.
}
\end{figure}
Equation~(\ref{inelastic}), $\delta I_{\textrm{1BA}}$, represents the inelastic process accompaying one vibron emission [Fig.~\ref{1BA}], 
which is essentially the same with the first Born approximation (1BA) in scattering theory~\cite{PRB2009Kristensen, PRB1970Davis}.
%
%
Equation~(\ref{elastic}), $\delta I_{\textrm{2BA}}$, correspond to the interference between the unperturbed state 
and the state perturbed by the second-order emission-reabsorption process, which can be obtained from the second Born approximation (2BA)~\cite{PRB1970Davis}. 
%
%
%
%
The first contribution in Eq.~(\ref{elastic}) is an one-electron scattering process and the corresponding schematic process is in Fig.~\ref{2BA} (a). 
The conducting electron initially occupying on the left scattering state $|\Phi_{Lm} (\varepsilon) \rangle$ 
is scattered off to the right scattering state $|\Phi_{Rn} (\varepsilon-\hbar\omega_\lambda) \rangle$, emitting one vibron $\hbar\omega_{\lambda}$. 
The corresponding transition amplitude is given by $\langle\Phi_{Rn}^{-}|\mathcal{M}^{\lambda}|\Phi_{Lm}\rangle$. 
After that, the same electron is excited to the right scattering state $|\Phi_{Rm} (\varepsilon) \rangle$ by absorbing the vibron that was emitted by the electron itself. 
$\langle\Phi_{Rm}|\mathcal{M}^{\lambda}|\Phi_{Rn}^{-}\rangle$ accounts for this second scattering event. 
%
%
In contrast, the second term in Eq.~(\ref{elastic}) involves two-electron scattering with a vibrational emission-reabsorption process. 
As shown in Fig.~\ref{2BA} (b), an electron on the left scattering state $|\Phi_{Ln} (\varepsilon+\hbar\omega_\lambda) \rangle$  first moves to the right scattering state $|\Phi_{Rm} (\varepsilon) \rangle$ by emitting one vibron $\hbar\omega_\lambda$.
Its transition amplitude is $\langle\Phi_{Rm}|\mathcal{M}^{\lambda}|\Phi_{Ln}^{+}\rangle$.
Subsequently, another electron at the left scattering state $|\Phi_{Lm} (\varepsilon) \rangle$ occupies the left scattering state $|\Phi_{Ln} (\varepsilon+\hbar\omega_\lambda) \rangle$ 
by absorbing the vibron previously emitted by the electron $1$ [Fig.~\ref{2BA} (b)].
This process gives $\langle\Phi_{Ln}^{+}|\mathcal{M}^{\lambda}|\Phi_{Lm}\rangle$. 

%
%
Note that the currents from these two elastic scattering processes have the opposite sign as seen in Eq.~(\ref{elastic}). 
This can be understood by the antisymmetry of two fermions under exchange operation.
When the two final states are compared in Figs.~\ref{2BA} (a) and (b) respectively, one can notice that they become identical by exchanging the electrons $1$ and $2$.
%
%
\begin{figure}[t]
\begin{center}
\includegraphics[width=0.9\columnwidth, clip=true]{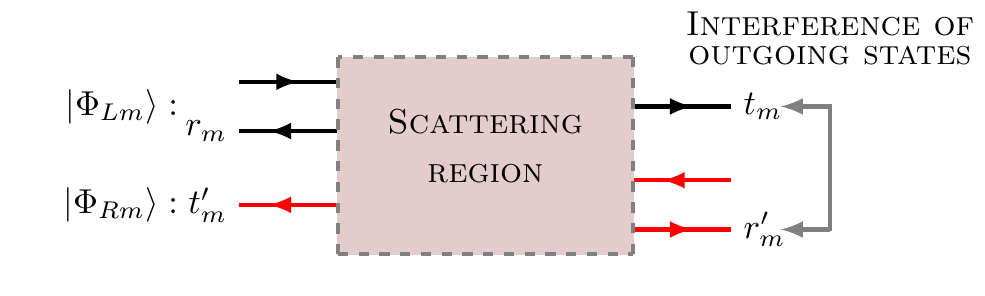} 
\end{center}
\caption{\label{prefactor} (Color online) For transmitted waves, the interference occurs between the outgoing waves of the unperturbed state (the left scattering state denoted by black arrows) $t_{m} | \Phi_{Rm}\rangle_{out}$ and
that of the second-order perturbed state (the right scattering state denoted by red arrows) $r^{\prime}_{m} | \Phi_{Rm}\rangle_{out}$ on the right electrode. The interference between these two outgoing waves
is proportional to $\textrm{Re}[r_{m}^\prime t_{m}]$.}
\end{figure}
Furthermore, the prefactor $r^{\prime}_{m} t_{m}^{*}$ in the elastic contribution [Eq.~(\ref{elastic})] implies that the elastic term originates from the interference between the zeroth-order state
and the second-order correction [Fig.~\ref{prefactor}]. While the outgoing state of the unperturbed state $|\Phi_{Lm} \rangle$ on the right side is $t_{m} | \Phi_{Rm}\rangle_{out}$, the outgoing state 
of the second-order perturbed state is $T_{(2)} r^{\prime}_{m} | \Phi_{Rm}\rangle_{out}$, where $T_{(2)}$ is the second-order transition coefficient discussed above. Explicitly, $T_{(2)}=\langle\Phi_{Rm}|\mathcal{M}^{\lambda}|\Phi_{Rn}^{-}\rangle\langle\Phi_{Rn}^{-}|\mathcal{M}^{\lambda}|\Phi_{Lm}\rangle$ for the one-electron 
elastic scattering process, and $T_{(2)}=\langle\Phi_{Rm}|\mathcal{M}^{\lambda}|\Phi_{Ln}^{+}\rangle\langle\Phi_{Ln}^{+}|\mathcal{M}^{\lambda}|\Phi_{Lm}\rangle$ for the two-electron process. The interference between the two outgoing states simply contributes to $\textrm{Re}\left[r^{\prime}_{m} t_{m}^{*} T_{(2)} \right]$, which can be seen in Eq.~(\ref{elastic}).
%
%
Unlike the previous study for a single channel model system with a very simple {\it el-vib} coupling~\cite{PRB1970Davis}, our formula in Eqs.~(\ref{inelastic}) and (\ref{elastic}) do not assume any specific form of Hamiltonians and generalizes to the multi-channel scattering system. We note that $\delta I_{1BA} + \delta I_{2BA}$ for a one-dimensional square-well potential with a single vibrational scatterer reduces to the current correction based on the Born series expansion~\cite{PRB1970Davis}. 

%
%
Equations~(\ref{I1}) and (\ref{I2}) in the NEGF formalism have been interpreted as inelastic and elastic processes respectively in the previous literatures~\cite{PRB2006Vega, PRL2008Paulsson, JPhysC1972Caroli, PRL2000Lorente, PRB2003Mii, JChP2004Galperin, PRB2005Viljas, SurfSci2007Ueba}. 
This interpretation has been made by using some heuristic arguments, for example, by indirectly inferring from Fermi distribution factors~\cite{PRB2005Viljas}. 
However, when the explicit expression of the {\it el-vib} self-energy is considered~\cite{PRB2007Frederiksen_b}, it is shown that Eqs. (\ref{I1}) and (\ref{I2}) have the same Fermi distribution factors like $f_{L}(1-f_{R}^{\pm})$ (see Appendix~\ref{Fermi_factors}). Therefore, one cannot distinguish which one is elastic or inelastic only by inspecting Fermi distribution factors. In contrast, we have specified Eqs.~(\ref{inelastic}) and (\ref{elastic}) as inelastic and elastic by directly keeping track of all the charge transitions between states based on a scattering theory.

\begin{figure}[t]
\begin{center}
\includegraphics[width=0.8\columnwidth, clip=true]{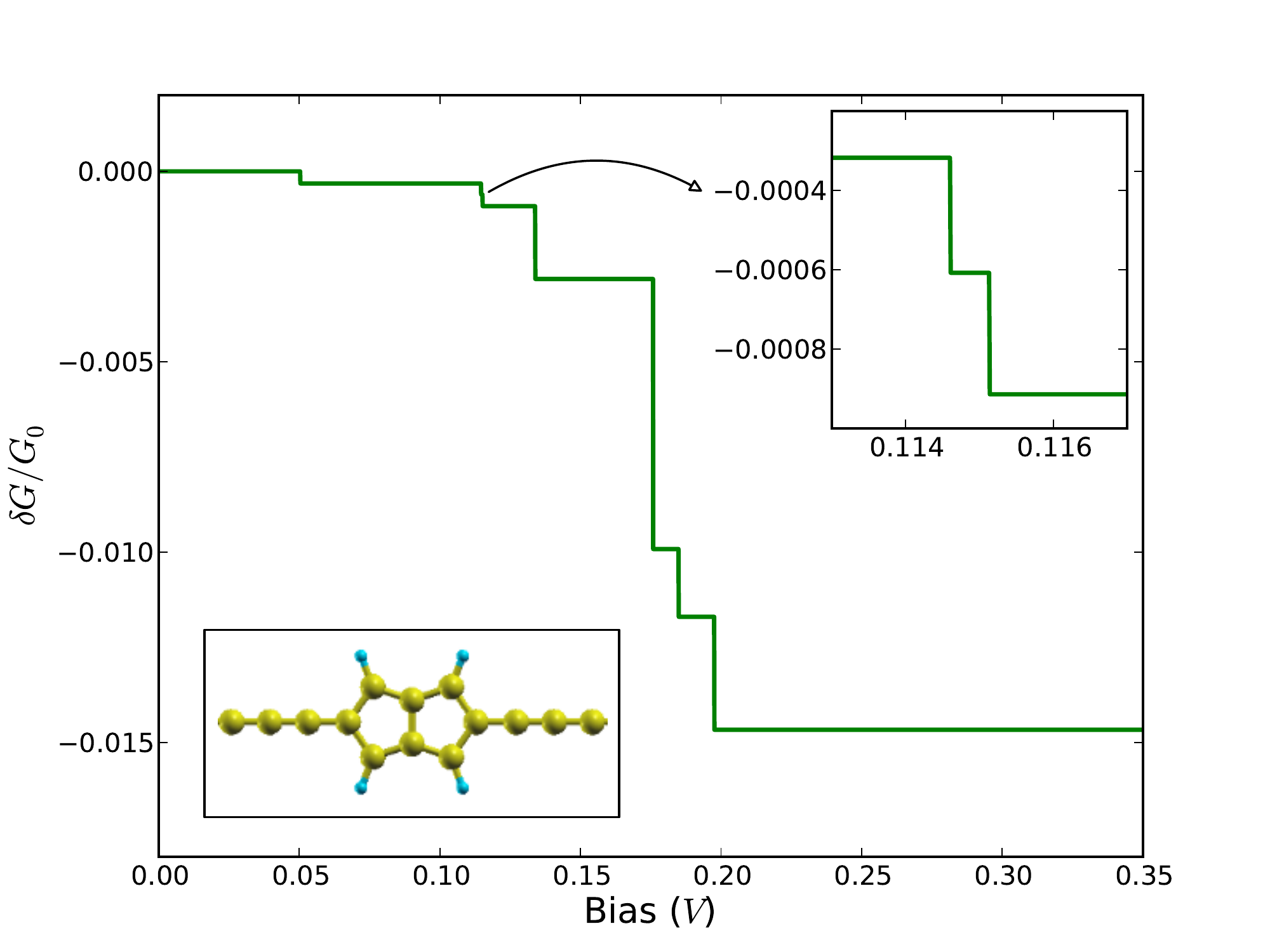} 
\end{center}
\caption{\label{cc-pentalene} (Color online) Differential conductance renormalized by $G_{0}=2e^2/h$ for a pentalene molecule coupled to monoatomic carbon chains from DFT calculation. 
Atomic configuration is indicated in the bottom left inset. 
Among $35$ vibrational modes, seven modes $\lambda=7, 18, 19, 21, 28, 29$, and $31$ lead to conductance steps from left to right.
The top right inset shows the second and third conductance steps corresponding to the modes $18$ and $19$, which are almost degenerate.
Corresponding vibrational energies are $\hbar\omega_{\lambda}=50.3, 114.6, 115.1, 133.8, 175.8, 184.9$, and $197.6$ meV respectively. 
}
\end{figure}

To illustrate, we perform a density functional theory (DFT) calculation on a pentalene molecule connected to monoatomic carbon chains~\cite{arXiv2012Sejoong}.
Seven vibrational modes of the pentalene lead to conductance steps as seen in Fig. \ref{cc-pentalene}.
Table \ref{table_cond} shows changes in the differential conductance $\delta G^{\lambda}$ when $eV=\hbar\omega_{\lambda}$, and separate contributions to $\delta G^{\lambda}$ for the seven active modes ($\lambda = 1, \cdots, 7$). 
Here, $\delta G^{\lambda}_{1}$ and $\delta G^{\lambda}_{2}$ are from the NEGF formalism for inelastic [Eq. (\ref{I1})] and elastic [Eq. (\ref{I2})] scattering processes 
while $\delta G^{\lambda}_{\textrm{1BA}}$ and $\delta G^{\lambda}_{\textrm{2BA}}$ are from our present scattering theory for inelastic [Eq. (\ref{inelastic})] and elastic [Eq. (\ref{elastic})] ones. 
Note that $\delta G^{\lambda} = \delta G_{1}^{\lambda} + \delta G_{2}^{\lambda} = \delta G_{\textrm{1BA}}^{\lambda} + \delta G_{\textrm{2BA}}^{\lambda}$ and that $\delta G_{1}^{\lambda} \neq \delta G_{\textrm{1BA}}^{\lambda}$ and  $\delta G_{2}^{\lambda} \neq \delta G_{\textrm{2BA}}^{\lambda}$. 
This indicates the contrast between the two interpretations. 

\begin{table}[t]
\caption{Numerical data for condutance changes $\delta G^{\lambda}$ at threshold bias voltage $eV=\hbar\omega_{\lambda}$ for modes of $\lambda$'s shown in Fig. \ref{cc-pentalene}.
$\delta G^{\lambda}$, $\delta G_{1}^{\lambda}$, $\delta G_{2}^{\lambda}$, $\delta G_{1BA}^{\lambda}$, and $\delta G_{2BA}^{\lambda}$ are written in units of $G_{0} \times 10^{-3}$.}
\label{table_cond}
\begin{tabular*}{0.48\textwidth}{@{\extracolsep{\fill}}c|r r r r r}
\hline 
mode ($\lambda$) & $\delta G^{\lambda}$ & $\delta G_{1}^{\lambda}$ & $\delta G_{2}^{\lambda}$ & $\delta G_{1BA}^{\lambda}$ & $\delta G_{2BA}^{\lambda}$\tabularnewline
\hline 
\hline 
 7 & $-0.317$ & 0.000 & $-0.317$ & $-0.270$ & $-0.047$  \\
18 & $-0.290$ & 0.000 & $-0.290$ & $-0.247$ & $-0.043$  \\
19 & $-0.306$ & 2.088 & $-2.394$ & $-0.306$ &    0.000  \\
21 & $-1.911$ & 0.000 & $-1.911$ & $-1.633$ & $-0.278$  \\
28 & $-7.092$ & 0.000 & $-7.092$ & $-6.059$ & $-1.033$  \\
29 & $-1.779$ & 0.300 & $-2.079$ & $-1.779$ &    0.000  \\
31 & $-2.968$ & 0.000 & $-2.968$ & $-2.535$ & $-0.433$  \\
\hline 
\end{tabular*}
\end{table}

\section{Discussion}\label{discussion}
\subsection{Single-channel case}
When our theory is applied to single-channel systems, one can derive a general formula for the conductance variation, 
in which the inelastic (1BA) and elastic (2BA) corrections are treated on an equal footing. 
For the case where the density of states of the system is slowly varying over a few vibrational energies around the Fermi energy $\varepsilon_{F}$,
one may use the following approximations: $G^{r}_{0}(\varepsilon) \approx G^{r}_{0}(\varepsilon_{F})$ and $\Gamma^{r}_{\alpha}(\varepsilon) \approx \Gamma^{r}_{\alpha}(\varepsilon_{F})$~\cite{PRB2005Paulsson}. 
Transmission and reflection coefficients, and scattering states may be replaced by those at $\varepsilon_{F}$. 
Then, using Eqs. (\ref{inelastic}) and (\ref{elastic}), the differential conductance has two components, $\delta  g= \delta g_{\textrm{1BA}} + \delta g_{\textrm{1BA}}$, where
\begin{eqnarray}
\label{G1} \delta g_{\textrm{1BA}} &=& \left(2\pi\right)^{2}\sideset{}{^\prime}\sum_{\lambda} \left(1-2\mathcal{T}\right)\left| \mathbb{M}^{\lambda}_{RL} \right|^{2} \\
\label{G2} \delta g_{\textrm{2BA}} &=& \left(2\pi\right)^{2}\sideset{}{^\prime}\sum_{\lambda} \textrm{Re}\left[ r^{\prime}t^{*} \mathbb{M}^{\lambda}_{RL} \left( \mathbb{M}^{\lambda}_{LL} - \mathbb{M}^{\lambda}_{RR} \right) \right].
\end{eqnarray} 
Here $\delta g_{\textrm{1BA(2BA)}}\equiv \delta G_{\textrm{1BA(2BA)}}/G_0$, and $\mathbb{M}^{\lambda}_{\alpha \beta} \equiv \langle \Phi_{\alpha} | \mathcal{M}^{\lambda} | \Phi_{\beta} \rangle$. 
$\sum_{\lambda}^\prime$ means that when it is summed over vibrational modes $\lambda$, the step function $\theta \left(\left|eV \right|-\hbar\omega_{\lambda} \right)$ is multiplied, {\it i.e.}, $\sum_{\lambda}^\prime \equiv \sum_{\lambda} \theta \left(\left|eV \right|-\hbar\omega_{\lambda} \right)$.
The inelastic contribution, which is proportional to $1-2\mathcal{T}$ in Eq. (\ref{G1}), changes its sign at $\mathcal{T}=0.5$. Then, when the elastic interference term [Eq. (\ref{G2})] is included as shown, the crossover transmission $\mathcal{T}_{\textrm{cr}}$
can deviate from $0.5$. In fact, using Eqs. (\ref{identity2-1}) and (\ref{identity2-2}), Eqs. (\ref{G1}) and (\ref{G2}) can be concisely expressed as follows:
\begin{equation}
\label{conductance2_single_channel}
\delta g =  \left( 2\pi \right)^2 \sideset{}{^\prime}\sum_{\lambda} \left(1 - 2\mathcal{T} - 2 \mathcal{R} \cos^{2} \theta_{\lambda} \right) \left| \mathbb{M}^{\lambda}_{RL} \right|^2 ,
\end{equation}
where $\theta_{\lambda} = \arg \left[r^{\prime} t^{*} \langle \Phi_{R} | \mathcal{M}^{\lambda} | \Phi_{L} \rangle \right]$. 
Equation (\ref{conductance2_single_channel}) has a form of Fermi's golden rule, which is proportional to $\left|\langle \Phi_{R} | \mathcal{M}^{\lambda} | \Phi_{L} \rangle \right|^2$. 
The elastic interference given by $-2 \mathcal{R} \cos^{2} \theta_{\lambda}$ is always negative. 
Since $\cos^{2} \theta_{\lambda}$ depends on the elastic transmission $\mathcal{T}$, one cannot obtain an analytic expression for the crossver transmission $\mathcal{T}_{\textrm{cr}}$.
However, it is obvious that the transition transmission $\mathcal{T}_{\textrm{cr}}$ is smaller than $0.5$, if the crossover occurs. 
When the elastic correction vanishes, Eq. (\ref{conductance2_single_channel}) recovers the $0.5$ rule. 

\subsubsection{Mirror symmetry}
%
%
For systems with some particular symmetries, one may calculate the elastic correction $-2\mathcal{R} \cos^{2} \theta_{\lambda}$ analytically.  
For example, let us consider mirror-symmetric systems along the transport direction, where $\cos^{2} \theta_{\lambda}$ is either $0$ or $1$ as shown below.  
Under the mirror reflection operator $\mathbf{R}$, left and right scattering states are related as follows: $\mathbf{R} | \Phi_{L,R} \rangle = |\Phi_{R,L} \rangle$.
For vibrational modes of the even mirror-reflection symmetry, which satisfies $\mathbf{R} \mathcal{M}^{\lambda}_{\textrm{even}} \mathbf{R}^{\dagger} = \mathcal{M}_{\textrm{even}}^{\lambda}$,
it can be shown that 
\begin{eqnarray}
\label{even-parity}
\delta g &=& (2\pi)^2 \sideset{}{^\prime}\sum_{\lambda} \left( 1 - 2 \mathcal{T} \right) \left| \langle \Phi_{R} | \mathcal{M}_{\textrm{even}}^{\lambda} | \Phi_{L} \rangle \right|^2,
\end{eqnarray}
since $\langle \Phi_{L} | \mathcal{M}_{\textrm{even}}^{\lambda} | \Phi_{L} \rangle = \langle \Phi_{R} | \mathcal{M}_{\textrm{even}}^{\lambda} | \Phi_{R} \rangle$.
Thus the $0.5$ rule holds for even mirror-symmetric vibrational modes. 
In contrast, for odd mirror-symmetric modes where $\mathbf{R} \mathcal{M}_{\textrm{odd}}^{\lambda} \mathbf{R}^{\dagger} = -\mathcal{M}_{\textrm{odd}}^{\lambda}$, 
the elastic interference term is reduced to be $-2\mathcal{R} \left| \langle \Phi_{R} | \mathcal{M}_{\textrm{odd}}^{\lambda} | \Phi_{L} \rangle \right|^2 $, 
or equivalently $\cos^{2} \theta_{\lambda} = 1$. Combining with the inelastic term, the differential conductance step becomes
\begin{equation}
\label{odd-parity}
\delta g = - (2\pi)^2 \sideset{}{^\prime}\sum_{\lambda} \left| \langle \Phi_{R} | \mathcal{M}_{\textrm{odd}}^{\lambda} | \Phi_{L} \rangle \right|^2.
\end{equation}
It means that odd mirror-symmetric vibrational modes always lead to downward conductance steps. 
Thus there is no crossover between upward and downward steps for these modes. 
These different behaviors of even and odd vibrational modes are already reported in the DFT calculation for conductance steps of a mirror-symmetric gold atomic junction~\cite{PRB2007Frederiksen_b}
and the study on a tight-binding model of a mirror-symmetric atomic chain~\cite{PRB2012Avriller}, which can readily be explained by our present theory. 

\begin{figure}[t]
\begin{center}
\includegraphics[width=0.9\columnwidth, clip=true]{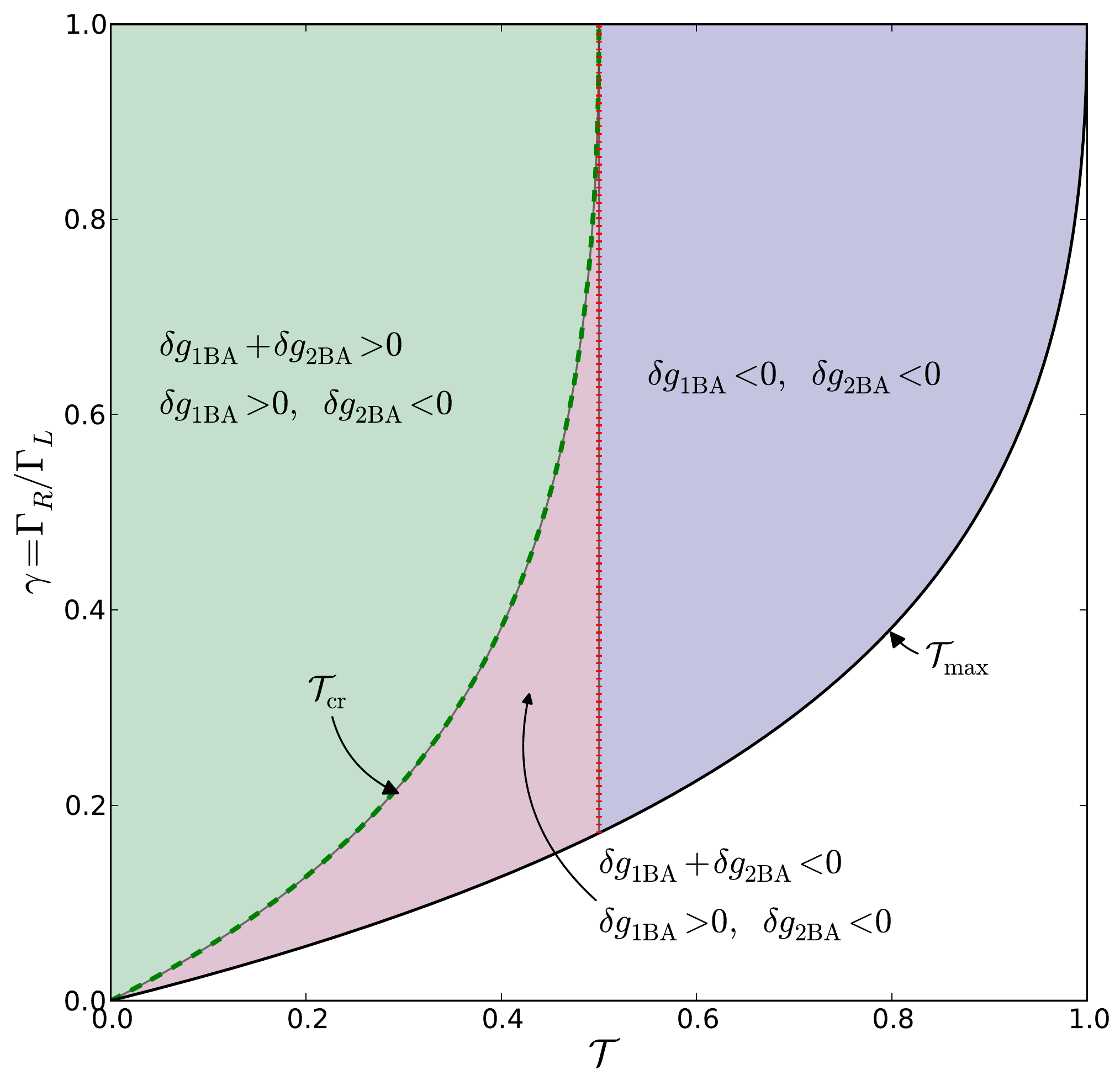} 
\end{center}
\caption{\label{single-level} Phase diagram of conductance variations due to inelastic scattering for the single-level model. 
Depending on signs of $\gamma_{\textrm{inel}}$ and $\gamma_{\textrm{inel}} + \gamma_{\textrm{el}}$, there are three regions indicated by green, red and blue colors. 
the conductance increases ($\gamma_{\textrm{inel}} + \gamma_{\textrm{el}} > 0$) in the green region, and it decreases ($\gamma_{\textrm{inel}} + \gamma_{\textrm{el}} < 0$) in the red and blue regions. 
Two phases are separated by $\mathcal{T}_{\textrm{cr}}$. In the red region, $\gamma_{\textrm{inel}} > 0$ and $\gamma_{\textrm{el}} < 0$, while both $\gamma_{\textrm{inel}}$ and $\gamma_{\textrm{el}}$
are negative in the blue region.}
\end{figure}
\subsubsection{Single-level Model}
In fact, the single-level model symmetrically connected to electrodes is regarded as the even mirror-reflection system. 
Considering the single-level model with a single vibration scatterer,
the inelastic and elastic contributions in our scattering description are
\begin{eqnarray}
\label{single_level_inelastic}\delta g_{\textrm{1BA}} &=&  \left| \frac{\mathcal{M}}{\Gamma_{L}} \right|^{2} \frac{\left(1-2\mathcal{T}\right)\mathcal{T}^2}{\gamma} \\
\label{single-level_elastic}  \delta g_{\textrm{2BA}} &=&  -\frac{1}{2} \left| \frac{\mathcal{M}}{\Gamma_{L}} \right|^{2} \frac{\left(1-\gamma \right)^2 }{\gamma^2} \mathcal{T}^{3},
\end{eqnarray}
where $\gamma = \Gamma_{R} / \Gamma_{L}$. 
A bare transmission $\mathcal{T}$ is given by 
\begin{equation}
\mathcal{T} = \frac{\Gamma_{L} \Gamma_{R}}{\left(\varepsilon_{F} - \varepsilon_{0} \right)^2 + \frac{\left(\Gamma_{L} + \Gamma_{R}\right)^2}{4}}.
\end{equation}

When the system is symmetrically coupled to electrodes, the elastic correction $\delta g_{\textrm{2BA}}$ vanishes 
and a crossover between an increase and a decrease in conductance is determined solely by the inelastic contribution $\delta g_{\textrm{1BA}}$. 
The inelastic contribution $\delta g_{\textrm{1BA}}$ [Eq.~(\ref{single_level_inelastic})] clearly shows the $0.5$ crossover. 
When the mirror symmetry is broken, {\it i.e.}, $\gamma \neq 1$, the elastic correction has a nonzero negative value as shown in Eq.~(\ref{single-level_elastic}),
and thus the crossover transmission $\mathcal{T}_{\textrm{cr}}$ is smaller than $0.5$. 
It is analytically shown that $\mathcal{T}_{\textrm{cr}} = 2 \gamma / \left(1 + \gamma \right)^2$, which is smaller than $0.5$ when $\gamma < 1$.

Reference~\onlinecite{PRL2008Paulsson} provides a phase diagram for the conductance variation of the single-level model in a space $\left(\mathcal{T}, \gamma \right)$.
The phase diagram can be re-drawn in Fig.~\ref{single-level} by highlighting inelastic and elastic contributions and their interplay. 
Depending on signs of $\delta g_{\textrm{1BA}}$ and $\delta g_{\textrm{1BA}} + \delta g_{\textrm{2BA}}$, there are three regions indicated in Fig.~\ref{single-level}. 
If the elastic contribution is neglected, the crossover is dertermined by the inelastic correction, and the phase boundary is given by $\mathcal{T}=0.5$ (red dotted line in Fig.~\ref{single-level}).
There, however, is a nonzero elastic correction in general except for $\gamma=1$, and this negative elastic correction shifts the phase boundary to $\mathcal{T}_{\textrm{cr}} = 2 \gamma / \left(1 + \gamma \right)^2 \leq 0.5$
(green dashed line in Fig.~\ref{single-level}). This result is consistent with Eq.~(\ref{conductance2_single_channel}), which holds not only to the single-level model, but also to any other single-channel system. 

\subsection{Multichannel Systems and Propensity Rules}
%
%
Next we remark on propensity rules \cite{PRL2008Paulsson, PRB2007Gagliardi, JChemPhys2006Troisi}. The conductance step for multi-channel systems is proportional to $\left| \langle \Phi_{Rn}|  \mathcal{M}^{\lambda} | \Phi_{Lm} \rangle \right|$ as seen in Eqs.~(\ref{inelastic}) and (\ref{elastic}). Therefore, investigating symmetry properties of scattering states $|\Phi_{Lm}\rangle$ and $|\Phi_{Rn}\rangle$, and {\it el-vib} interactions $\mathcal{M}^{\lambda}$, 
one can distinguish symmetry-prohibited transitions between scattering states as done in Ref.~\onlinecite{PRL2008Paulsson}. However, the height of the conductance step cannot be approximated by the ansatz of Ref.~\onlinecite{PRL2008Paulsson} as discussed in the single-channel case. Instead, 
Eqs.~({\ref{inelastic}) and ({\ref{elastic}) can be used to quantitatively analyze the height of the conductance step, for example, contributions of intra-channel and inter-channel scatterings to the height.

\section{conclusion} \label{conclusion}
We have presented the scattering-state description for inelastic transport, which is obtained by converting NEGFs to scattering states. Our description takes into account two competing scattering processes, elastic and inelastic ones, one of which is missed in some other scattering theory approaches. Based on scattering states and charge transitions among them, we have clarified elastic and inelastic scattering processes leading to the conductance variations. Importantly, the specification on the two competing contributions enables to understand the crossover of the conductance step in general situations. When applying our result to single-channel systems, we have shown that the crossover transmission is generally smaller than $0.5$, or even the crossover does not appear since the elastic contribution is negative. For multichannel systems, our expression can be useful to investigate how scattering processes between transmission channels lead to conductance changes.

\section{acknowledgement}
We thank Hyun-Woo Lee, Mahn-Soo Choi, and Myung-Joong Hwang for helpful discussions. Young-Woo Son was supported by the NRF grant funded by Korean government MEST (Quantum Metamaterials research center, No. R11-2008-053-01002-0 and the Center for Advanced Soft Electronics 2011-0031640). Computation was supported by the CAC of KIAS.
For DFT calculations, we have used \textsc{QUANTUM-ESPRESSO}~\cite{JPhys2009Giannozzi} and \textsc{WANNIER90}~\cite{ComputPhysCommun2008Mostofi}.

\appendix
\section{Derivation of Eqs.~(\ref{I1}) and (\ref{I2}) and Fermi distribution factors} \label{Fermi_factors}
In the NEGF formalism~\cite{JChP2004Galperin, PRB2005Viljas, PRB2007Frederiksen_a}, the current measured at the left electrode to the second order of {\it el-vib} couplings is 
\begin{equation}
I_{\alpha} = I_{\textrm{el}}^{(0)} + I_{\textrm{el}}^{(2)} + I_{\textrm{inel}}^{(2)},
\end{equation}
where
\begin{widetext}
\begin{eqnarray}
\label{Landauer} I_{\textrm{el}}^{(0)} &=& \frac{2e}{h} \int^{\infty}_{-\infty} d\varepsilon \textrm{Tr} \left[\Gamma_{L} G_{0}^{r} \Gamma_{R} G^{a}_{0} \right] \left[f_{L} - f_{R} \right] \\
\label{inel}     I_{\textrm{inel}}^{(2)} &=& \frac{2e}{h} \int^{\infty}_{-\infty} d\varepsilon i\textrm{Tr} \left[\Gamma_{L} G^{r}_{0} \Sigma^{>,(2)}_{el-vib} G_{0}^{a} f_{L} - \Gamma_{L} G^{r}_{0} \Sigma^{<,(2)}_{el-vib} G_{0}^{a} (f_{L}-1) \right] \\
\label{el}       I_{\textrm{el}}^{(2)} &=& \frac{2e}{h} \int^{\infty}_{-\infty} d\varepsilon 2 \textrm{Re} \textrm{Tr} \left[\Gamma_{L} G_{0}^{r} \Sigma^{r,(2)}_{el-vib} G^{r}_{0} \Gamma_{R} G^{a}_{0} \right] \left[f_{L} - f_{R} \right].
\end{eqnarray} 
Here $\Sigma^{r,(2)}_{el-vib}$, $\Sigma^{<,(2)}_{el-vib}$, and $\Sigma^{>,(2)}_{el-vib}$ denote the retarded, lesser, and greater electron-vibration self-energies respectively~\cite{PRB2007Frederiksen_a}. 
Equations~(\ref{I1}) and (\ref{I2}) leading to conductance steps are obtained from Eqs. (\ref{inel}) and (\ref{el}) respectively.
Equation (\ref{Landauer}) is the well-known Landauer formula for the elastic current without electron-vibration scattering. 
In the literatures~\cite{JChP2004Galperin, PRB2005Viljas}, Eq. (\ref{el}) is identified as the elastic correction, because it has the same Fermi distribution factor $(f_{L}-f_{R})$ as the Landauer formula does.
Equation~(\ref{inel}) is specified as the inelastic current. 

However, when the explicit expressions of $\Sigma^{r,(2)}_{el-vib}$, $\Sigma^{<,(2)}_{el-vib}$, and $\Sigma^{>,(2)}_{el-vib}$ are used~\cite{PRB2007Frederiksen_a}, 
$\delta I_{1}$ and $\delta I_{2}$ corresponding to Eqs.~(\ref{I1}) and (\ref{I2}) are given as follows:
\begin{eqnarray}
\label{I1_original}\delta I_{1} &=& \sum_{\beta = L, R}\sum_{\lambda} \frac{2e}{h} \int^{\infty}_{-\infty} d\varepsilon \textrm{Im}\textrm{Tr}\left[\Gamma_{L} G^{r}_{0} \mathcal{M}^{\lambda} A^{+}_{\beta} \mathcal{M}^{\lambda} A_{R}\right]  N_{\lambda} \left[ f_{L}(1-f_{\beta}^{+}) - f_{R}(1-f_{\beta}^{+}) \right] \nonumber \\
             & & \qquad\qquad\qquad\qquad + \textrm{Im}\textrm{Tr}\left[\Gamma_{L} G^{r}_{0} \mathcal{M}^{\lambda} A^{+}_{\beta} \mathcal{M}^{\lambda} A_{R}\right] (N_{\lambda} + 1) \left[ f_{\beta}^{+}(1 - f_{R}) - f_{\beta}^{+}(1-f_{L})  \right] \nonumber \\
             & & \qquad\qquad\qquad\qquad + \textrm{Im}\textrm{Tr}\left[\Gamma_{L} G^{r}_{0} \mathcal{M}^{\lambda} A^{-}_{\beta} \mathcal{M}^{\lambda} A_{R}\right] N_{\lambda} \left[ f_{\beta}^{-}(1 - f_{R}) - f_{\beta}^{-}(1-f_{L})   \right] \nonumber \\
             & & \qquad\qquad\qquad\qquad + \textrm{Im}\textrm{Tr}\left[\Gamma_{L} G^{r}_{0} \mathcal{M}^{\lambda} A^{-}_{\beta} \mathcal{M}^{\lambda} A_{R}\right] (N_{\lambda} + 1) \left[ f_{L}(1-f_{\beta}^{-}) - f_{R}(1-f_{\beta}^{-}) \right] \\
\label{I2_original}\delta I_{2} &=& \sum_{\beta = L, R}\sum_{\lambda} \frac{2e}{h} \int^{\infty}_{-\infty} d\varepsilon \textrm{Tr}\left[G^{a}_{0}\Gamma_{L}G^{r}_{0}\mathcal{M}^{\lambda}A_{\beta}^{-}\mathcal{M}^{\lambda}\right] \left[ (N_{\lambda}+1) f_{L} (1-f_{\beta}^{-}) - N_{\lambda}f_{\beta}^{-}(1-f_{L}) \right] \nonumber \\
             & & \qquad\qquad\qquad\qquad + \textrm{Tr}\left[G^{a}_{0}\Gamma_{L}G^{r}_{0}\mathcal{M}^{\lambda}A_{\beta}^{+}\mathcal{M}^{\lambda}\right] \left[ N_{\lambda} f_{L} (1 - f_{\beta}^{+}) - (N_{\lambda}+1) f_{\beta}^{+} (1-f_{L}) \right],
\end{eqnarray} 
where $N_{\lambda}$ denotes the population for the vibratioal mode $\lambda$. Note that in a zero temperature limit and a regime of the externally thermalization considered in the paper, {\it i.e.}, $N_{\lambda} \approx 0$, Eqs.~(\ref{I1_original}) and (\ref{I2_original}) are reduced to  Eqs.~(\ref{I1}) and (\ref{I2}). As clearly shown above, it turns out that both Eqs.~(\ref{I1_original}) and (\ref{I2_original}) include only the Fermi distribution factors like $f_{\alpha} (1-f_{\beta}^\pm)$ rather than $(f_{L}-f_{R})$. 

\section{Derivation of Eqs.~(\ref{inelastic}) and (\ref{elastic})} \label{derivation}
Equations.~(\ref{I1}) and (\ref{I2}) are explicitly expressed in terms of scattering states $\left\{ |\Phi_{\alpha m} \rangle \right\}$ and scattering matrices 
$\mathcal{S}_{m}=\left(\begin{smallmatrix} r_{m} & t_{m}^{\prime}\\ t_{m} & r_{m}^{\prime}\end{smallmatrix}\right)$ as follows:
\begin{eqnarray}
\label{I1-2}
\delta I_{1} = \frac{2e}{h}\left(2\pi\right)^{2} \sum_{\lambda} \sum_{m,n} \int_{\mu_{R}+\hbar\omega_{\lambda}}^{\mu_{L}} d\varepsilon && \left[ \mathcal{R}_{m} \left| \langle\Phi_{Rn}^{-}| \mathcal{M}^{\lambda} |\Phi_{Lm}\rangle \right|^{2} 
                 + \mathcal{T}_{m} \left| \langle\Phi_{Rn}^{-}| \mathcal{M}^{\lambda} |\Phi_{Rm}\rangle \right|^{2} \right. \nonumber \\
             & & \left. - 2\textrm{Re}\left\{ r^{\prime}_{m}t^{*}_{m} \langle\Phi_{Rm}| \mathcal{M}^{\lambda} |\Phi_{Rn}^{-}\rangle \langle\Phi_{Rn}^{-}| \mathcal{M}^{\lambda} |\Phi_{Lm}\rangle \right\} \right],
\end{eqnarray} 
\begin{eqnarray}
\label{I2-2}
\delta I_{2} &=& \frac{2e}{h}\left(2\pi\right)^{2} \sum_{\lambda} \sum_{m,n} \int_{\mu_{R}}^{\mu_{L}-\hbar\omega_{\lambda}} d\varepsilon \left[ -\mathcal{T}_{n} \left| \langle\Phi_{Rn}|\mathcal{M}^{\lambda} |\Phi_{Lm}^{+}\rangle \right|^{2} + \textrm{Re} \left\{ r_{n}^{\prime} t_{n}^{*} \langle\Phi_{Rn}|\mathcal{M}^{\lambda}|\Phi_{Lm}^{+}\rangle \langle\Phi_{Lm}^{+}|\mathcal{M}^{\lambda}|\Phi_{Ln}\rangle\right\} \right]  \nonumber \\
             & & +\frac{2e}{h}\left(2\pi\right)^{2} \sum_{\lambda} \sum_{m,n} \int_{\mu_{R}+\hbar\omega_{\lambda}}^{\mu_{L}} d\varepsilon \left[-\mathcal{T}_{m}\left| \langle\Phi_{Rn}^{-}| \mathcal{M}^{\lambda} |\Phi_{Rm}\rangle \right|^{2} + \textrm{Re}\left\{ r_{m}^{\prime} t_{m}^{*} \langle\Phi_{Rm}| \mathcal{M}^{\lambda} |\Phi_{Rn}^{-}\rangle \langle\Phi_{Rn}^{-}| \mathcal{M}^{\lambda} |\Phi_{Lm}\rangle \right\} \right],
\end{eqnarray}
where superscripts $\pm$ indicate that energy argument is $\varepsilon \pm \hbar \omega_{\lambda}$ and without $\pm$, the argument is $\varepsilon$. 
Collecting terms in a form of $\left| \langle\Phi_{\alpha m}| \mathcal{M}^{\lambda} |\Phi_{\beta n} \rangle \right|^{2}$ from Eqs. (\ref{I1-2}) and (\ref{I2-2}), 
one can obtain $\delta I_{\textrm{1BA}}$, 
\begin{eqnarray}
\delta I_{\textrm{1BA}} &=& \frac{2e}{h}\left(2\pi\right)^{2} \sum_{\lambda} \sum_{m,n} \left[ \int_{\mu_{R}+\hbar\omega_{\lambda}}^{\mu_{L}} d\varepsilon \mathcal{R}_{m} \left| \langle\Phi_{Rn}^{-}| \mathcal{M}^{\lambda} |\Phi_{Lm}\rangle \right|^{2} 
                   - \int_{\mu_{R}}^{\mu_{L}-\hbar\omega_{\lambda}} d\varepsilon \mathcal{T}_{n} \left| \langle\Phi_{Rn}|\mathcal{M}^{\lambda} |\Phi_{Lm}^{+}\rangle \right|^{2} \right] \nonumber \\
               &=& \frac{2e}{h}\left(2\pi\right)^{2} \sum_{\lambda} \sum_{m,n} \int_{\mu_{R}+\hbar\omega_{\lambda}}^{\mu_{L}} d\varepsilon \left(\mathcal{R}_{n}^{-} - \mathcal{T}_{m}\right) \left| \langle\Phi_{Rn}^{-}| \mathcal{M}^{\lambda} |\Phi_{Lm}\rangle \right|^{2}.
\end{eqnarray}
Similarly, a sum of terms of $\textrm{Re}\left\{ \cdots \right\}$ in Eqs. (\ref{I1}) and (\ref{I2}) is reduced to 
\begin{eqnarray}
\delta I_{\textrm{2BA}} &=& -\frac{2e}{h}\left(2\pi\right)^{2}\sum_{\lambda}\sum_{m,n}\int_{\mu_{R}+\hbar\omega_{\lambda}}^{\mu_{L}}d\varepsilon \textrm{Re}\left[  r_{m}^{\prime}t_{m}^{*}\langle\Phi_{Rm}|\mathcal{M}^{\lambda}|\Phi_{Rn}^{-}\rangle\langle\Phi_{Rn}^{-}|\mathcal{M}^{\lambda}|\Phi_{Lm}\rangle\right]  \nonumber\\
                                        & & +\frac{2e}{h}\left(2\pi\right)^{2}\sum_{\lambda}\sum_{m,n}\int_{\mu_{R}}^{\mu_{L}-\hbar\omega_{\lambda}}d\varepsilon \textrm{Re}\left[r_{m}^{\prime}t_{m}^{*}\langle\Phi_{Rm}|\mathcal{M}^{\lambda}|\Phi_{Ln}^{+}\rangle\langle\Phi_{Ln}^{+}|\mathcal{M}^{\lambda}|\Phi_{Lm}\rangle\right].
\end{eqnarray}
As clearly seen above, $\delta I_{1} \neq \delta I_{\textrm{1BA}}$ and $\delta I_{2} \neq \delta I_{\textrm{2BA}}$. 
\end{widetext}

\end{document}